\documentclass[aip,pop]{revtex4-2}
\usepackage{amsmath}
\usepackage{amsfonts}
\usepackage{amssymb}
\usepackage{graphicx}

\def\fd#1#2{\frac{\delta #1}{\delta #2}}
\def\Sb#1#2{\overline{S}_{#1#2}}
\def\Ub#1#2{\overline{U}_{#1#2}}
\def\fm{f_{\mathrm m}}
\def\Ft{\widetilde{F}}

\begin{document}

\title{Hamiltonian fluid reduction of the 1.5D Vlasov-Maxwell equations}

\author{C. Chandre}
\email{cristel.chandre@univ-amu.fr}
\affiliation{CNRS, Aix Marseille Univ, Centrale Marseille, I2M, Marseille, France}

\author{B. A. Shadwick}
\email{shadwick@unl.edu}
\affiliation{University of Nebraska-Lincoln, Lincoln, NE, United States}

\begin{abstract}
We consider the Vlasov-Maxwell equations with one spatial direction and two momenta, one in the longitudinal direction and one in the transverse direction.  By solving the Jacobi identity, we derive
reduced Hamiltonian fluid models for the density, the fluid momenta and the second order moments, related to the pressure tensor.  We also provide the Casimir invariants of the reduced Poisson
bracket.  We show that the linearization of the equations of motion around homogeneous equilibria reproduces some essential feature of the kinetic model, the Weibel instability.
  
\end{abstract}

\maketitle

\section{Introduction}

The Vlasov--Maxwell system on a three dimensional phase space (one space and two momentum directions) is the minimal plasma configuration supporting electromagnetic modes and is thus of fundamental 
interest. We consider a system of $N$ charge particles of mass $m$ and charge $q$, described by a distribution function $f(z,p_x,p_z,t)$ and electromagnetic fields ${\bf E}=E_x(z,t)\,\hat{\bf x}+E_z(z,t)\,\hat{\bf
z}$ and ${\bf B}=B_y(z,t)\,\hat{\bf y}$.  The distribution function satisfies
$$
\int {\rm d}z {\rm d}p_x {\rm d}p_z\; f(z,p_x,p_z,t)=N,
$$
for all time. The dynamics is given by:
\begin{eqnarray}
&& \frac{\partial f}{\partial t}=-v_z\frac{\partial f}{\partial z}-q\left(E_x-\frac{v_z}{c} B_y\right)\frac{\partial f}{\partial p_x}-q\left(E_z+\frac{v_x}{c} B_y\right)\frac{\partial f}{\partial p_z},\label{eqn:VM15D_d}\\
&& \frac{\partial E_x}{\partial t}=-c\frac{\partial B_y}{\partial z}-4\pi q\int {\rm d}p_x {\rm d}p_z v_x f,\\
&& \frac{\partial E_z}{\partial t}=-4\pi q \int {\rm d}p_x {\rm d}p_z v_z f,\\
&& \frac{\partial B_y}{\partial t}=-c \frac{\partial E_x}{\partial z},\label{eqn:VM15D_f}
\end{eqnarray}
where $v_z$ and $v_x$ are the velocities in the $z$- (longitudinal) and $x$- (transverse) directions.  For instance, in the non-relativistic case, we have $p_z=m v_z$ and $p_x=m v_x$.  In what follows, we omit the implicit
time-dependence of the field variables $f$, $E_x$, $E_z$ and $B_y$, as is usual for dynamical variables. In the literature, this model is referred to as the 1.5D Vlasov-Maxwell system (see, e.g., 
Refs.~\onlinecite{Glassey90,Nguyen15,Glassey16} where the well-posedness of these equations is addressed).

There are systems where the entirety of the phase space data embodied in the distribution function is not relevant and a low-order moment description may be suitable.  For example, in a laser-driven 
plasma accelerator \cite{Esarey:2009sf} the phase velocity of the wakefield is typically well below the thermal velocity making wave-particle resonance unimportant to evolution of the laser field and 
the generation of plasma waves.  Nonetheless, it is of interest to capture the behavior beyond the cold fluid approximation,\cite{Shadwick05a} but numerical solutions of the Vlasov--Maxwell system is 
wholly impractical.  Fluid closures represent a practical means of incorporating this physics at a reasonable computational cost.\cite{Shadwick04b}

In addition, it is very convenient to work with fluid moments like the fluid density, fluid velocity and pressure tensor. These fluid variables are often more intuitive than their kinetic counterparts since they are functions in the configuration space rather than phase space. By taking moments of Eq.~(\ref{eqn:VM15D_d}), we notice that the equations of motion for the moments of order $K$ depends on the moment of order $K+1$.  From the Vlasov equation, we
construct an infinite ladder of fluid equations.  This ladder has to be truncated for practical purposes, and this truncation is a highly non-trivial problem.  The term at the origin of this conundrum
is the convective term $-v_z \partial f/\partial z$.

There are many ways of truncating this ladder.  Usually, it involves an ansatz for the distribution function (e.g., a Maxwellian~\cite{Shadwick04,Shadwick05,Shadwick12}, a bi-Maxwellian distribution
function~\cite{Goswami05} or a sum of Dirac functions~\cite{Fox09,Yuan11,Chalons12,Cheng14}), from which we compute the higher order moments as functions of the low-order moments.  Another way is to
include a dissipative term which emulates the transfers of energy from lower to higher order moments, as represented, e.g., by the growth rate of kinetic instabilities (see,
e.g., Ref.~\onlinecite{Hammett90}).

In this article, we follow the route initiated in Refs.~\onlinecite{Perin14,Perin15} for the one-dimensional Vlasov-Amp\`ere equations.  We require that the truncated system of equations preserves an
important property of the parent Vlasov-Maxwell equations, namely its Hamiltonian structure.  The Jacobi identity imposes numerous constraints on the truncated system which need to be satisfied to
properly define a Hamiltonian fluid model.  This procedure belongs to a class of methods, Hamiltonian reductions, used to reduce the set of variables of a Hamiltonian system while preserving the Hamiltonian structure.

In two phase-space dimensions (one spatial and one velocity), it was found in Refs.~\onlinecite{Yu00,Chesnokov12,Perin15wb} that the waterbag closures~\cite{Roberts67,Bertrand68,Berk70} correspond to a
Hamiltonian closure, the main reason being that the waterbag is an exact solution of the Vlasov equation.  However, this reduction is no longer possible in higher dimensions, where one has to deal
with at least two velocities.  Here we consider the minimal situation where the waterbag closure is not applicable to derive reduced Hamiltonian fluid models: one spatial direction and two velocities.
The parent system is the 1.5D Vlasov-Maxwell equations as given by Eqs.~(\ref{eqn:VM15D_d}-\ref{eqn:VM15D_f}).

In Sec.~\ref{sec:ham}, we provide the Hamiltonian structure of the parent model, the 1.5D Vlasov-Maxwell equations.  In Sec.~\ref{sec:hfr}, we define the reduced variables, i.e., the fluid density,
the two fluid momenta and the three components of the pressure tensor, and we provide the reduced bracket.  Then we detail and solve the constraints imposed on the third order moments by the Jacobi
identity in order to define reduced Hamiltonian fluid models.  In addition, we provide the Casimir invariants of the non-canonical Poisson bracket of these reduced systems.  In Sec.~\ref{sec:eqn}, we investigate the associated linearized system of equations around homogeneous equilibria. We show that by adjusting the free parameters of the reduced Hamiltonian fluid models, we can reproduce an essential result of the kinetic model, namely the instability of low-wavelength modes under temperature anisotropy.  

\section{Hamiltonian structure of the 1.5D Vlasov-Maxwell system}
\label{sec:ham}

The system of equations~(\ref{eqn:VM15D_d}-\ref{eqn:VM15D_f}) is Hamiltonian with a non-canonical Poisson bracket~\cite{Morrison80,Marsden82} (see also Refs.~\onlinecite{Morrison82,Morrison98} for an
introduction to non-canonical Hamiltonian systems in fluid and plasma physics).  The Poisson bracket is given by
\begin{eqnarray}
\{F,G\}&=& \int {\rm d}z {\rm d}p_x {\rm d}p_z\; f \left[\left(\frac{\partial}{\partial z}\fd Ff\right)\frac{\partial}{\partial p_z} \fd Gf-\left(\frac{\partial}{\partial p_z} \fd Ff\right)\frac{\partial}{\partial z} \fd Gf\right]\nonumber\\
&&{} + \frac{q}{c}\int {\rm d}z {\rm d}p_x {\rm d}p_z\; f B_y \left[\left(\frac{\partial}{\partial p_z} \fd Ff\right)\frac{\partial}{\partial p_x} \fd Gf-\left(\frac{\partial}{\partial p_x} \fd Ff\right)\frac{\partial}{\partial p_z} \fd Gf\right]\nonumber\\
&&{} + 4\pi q \int {\rm d}z {\rm d}p_x {\rm d}p_z\; f \left[\fd G{E_x}\frac{\partial}{\partial p_x} \fd Ff +\fd G{E_z}\frac{\partial}{\partial p_z} \fd Ff -\fd F{E_x} \frac{\partial}{\partial p_x} \fd Gf -\fd F{E_z}\frac{\partial}{\partial p_z} \fd Gf  \right]\nonumber\\
&&{} + 4\pi c\int {\rm d}z \left(\fd G{E_x} \frac{\partial}{\partial z} \fd F{B_y}-\fd F{E_x}\frac{\partial}{\partial z} \fd G{B_y}\right).\label{eqn:brackVM15D}
\end{eqnarray}
The equations of motion are obtained with $\dot{F}=\{F,H\}$.  In particular, the equations $\dot{f}=\{f,H\}$, $\dot{E}_x=\{E_x,H\}$, $\dot{E}_z=\{E_z,H\}$ and $\dot{B}_y=\{B_y,H\}$, with the
Hamiltonian
$$
H[f,E_x,E_z,B_y]=\int {\rm d}z {\rm d}p_x {\rm d}p_z\; f {\cal K}(p_x,p_z) +\int {\rm d}z\;\frac{E_x^2+E_z^2+B_y^2}{8\pi},
$$
where ${\cal K}(p_x,p_z)=(p_x^2+p_z^2)/(2m)$ in the non-relativistic case, and ${\cal K}(p_x,p_z)= mc^2[1+(p_x^2+p_z^2)/m^2 c^2]^{1/2}$ in the relativistic case, are identical to
Eqs.~(\ref{eqn:VM15D_d}-\ref{eqn:VM15D_f}).  The velocities $v_x$ and $v_z$ are given by $v_x=\partial {\cal K}/\partial p_x$ and $v_z=\partial {\cal K}/\partial p_z$.

We verify that this bracket is a Poisson bracket, i.e., it is bilinear, antisymmetric, satisfies the Leibniz rule and the Jacobi identity given by
$$
\{F_1,\{F_2,F_3\}\}+\{F_3,\{F_1,F_2\}\}+\{F_2,\{F_3,F_1\}\}=0,
$$   
for all observables $F_1$, $F_2$ and $F_3$. A direct proof of the Jacobi identity is possible following Refs.~\onlinecite{Morrison82,Morrison13}. A more straightforward verification  is obtained by using the canonical momenta 
\begin{eqnarray*}
&& \pi_x=p_x+\frac{q}{c}A_x(z),\\
&& \pi_z=p_z+\frac{q}{c}A_z(z),
\end{eqnarray*}
where $B_y=\partial A_x/\partial z$. Expressing the distribution function in these variables $\fm(z,\pi_x,\pi_z)=f(z,p_x,p_z)$, the following Poisson bracket
\begin{eqnarray*}
 \{F,G\} &=& \int {\rm d}z {\rm d}\pi_x {\rm d}\pi_z\;
\fm \left[\left(\frac{\partial}{\partial z}\fd F{\fm}\right)\frac{\partial}{\partial\pi_z}\fd G{\fm}
  - \left(\frac{\partial}{\partial\pi_z} \fd F{\fm}\right)  \frac{\partial}{\partial z} \fd G{\fm}\right]\\[2pt]
&&{} +4\pi c \int  {\rm d}z \left(\fd F{E_x} \fd G{A_x}+\fd F{E_z} \fd G{A_z}- \fd F{A_x} \fd G{E_x}- \fd F{A_z} \fd G{E_z}\right),
\end{eqnarray*}
reduces to the Poisson bracket~(\ref{eqn:brackVM15D}) in the Coulomb gauge $\partial A_z/\partial z=0$.
From the expression of the bracket in terms of the canonical momenta, it is easier to verify the Jacobi identity (see also Ref.~\onlinecite{Marsden82}).  As a consequence, the resulting system
(\ref{eqn:VM15D_d}--\ref{eqn:VM15D_f}) is a Hamiltonian system.

\section{Hamiltonian fluid reduction}
\label{sec:hfr}

The next step is to define fluid moments in the following way:
\begin{eqnarray*}
&& \rho=\int {\rm d}p_x {\rm d}p_z\ f,\\
&& P_x=\frac1\rho\int {\rm d}p_x {\rm d}p_z\ p_x\ f,\\
&& P_z=\frac1\rho\int {\rm d}p_x {\rm d}p_z\  p_z\ f,\\
&& S_{nk}=\frac{1}{\rho^{k+1}} \int {\rm d}p_x {\rm d}p_z\ (p_x-P_x)^n (p_z-P_z)^k\, f.
\end{eqnarray*}
We want to keep the second-order moments (related to the pressure tensor) as dynamical variables, i.e., we consider $\rho$, $P_x$, $P_z$, $S_{20}$, $S_{11}$ and $S_{02}$ as dynamical variables characterizing the distribution function $f$. We keep $E_x$, $E_z$ and $B_y$ as dynamical variables characterizing the electromagnetic field.  

\subsection{Expression of the bracket}
\label{Sec:brack}

We perform the reduction by considering the subset of observables which only depends on the first moments, i.e., the subset of
$\Ft[\rho(z),P_x(z),P_z(z),S_{20}(z),S_{11}(z),S_{02}(z),E_x(z),E_z(z),B_y(z)]$.  This subset is a priori not a Poisson subalgebra.
In order to compute the bracket, we use the following reduction:
\begin{multline*}
F[f(z,p_x,p_z),E_x(z),E_z(z),B_y(z)]\\=\Ft[\rho(z),P_x(z),P_z(z),S_{20}(z),S_{11}(z),S_{02}(z),E_x(z),E_z(z),B_y(z)],
\end{multline*}
and we obtain the following relation between the functional derivatives
\begin{eqnarray*}
\fd Ff&=&\fd\Ft\rho+\frac{p_x-P_x}{\rho}\fd\Ft{P_x}+\frac{p_z-P_z}{\rho}\fd\Ft{P_z}+\frac{(p_x-P_x)^2-S_{20}}{\rho}\fd\Ft{S_{20}},\\
&&{} +\frac{(p_x-P_x)(p_z-P_z)-2\rho S_{11}}{\rho^2}\fd\Ft{S_{11}}+\frac{(p_z-P_z)^2-3\rho^2 S_{02}}{\rho^3}\fd\Ft{S_{02}}\,.
\end{eqnarray*}
We insert this expression for the functional derivative into the bracket~(\ref{eqn:brackVM15D}), and we obtain the reduced bracket
\begin{eqnarray}
\{F,G\}&=& \int {\rm d}z \left[\left( \frac{\partial}{\partial z} \fd F\rho-4\pi q \fd F{E_z}\right)\fd G{P_z}-\fd F{P_z}\left( \frac{\partial}{\partial z} \fd G\rho - 4\pi q \fd G{E_z}\right)\right.\nonumber \\[2pt]
&&{} + 4\pi\left(c\frac{\partial}{\partial z}\fd F{B_y} + q \fd F{P_x}\right)\fd G{E_x}-4\pi \fd F{E_x}\left( c\frac{\partial}{\partial z} \fd G{B_y} + q \fd G{P_x}\right)\nonumber \\[2pt]
&&{} - \frac1\rho\left(\frac{q B_y}{c} + \frac{\partial P_x}{\partial z}\right)\left[\fd F{P_x}\fd G{P_z} - \fd F{P_z}\fd G{P_x} + 2\frac{S_{20}}\rho\left(\fd F{S_{20}}\fd G{S_{11}} -  \fd F{S_{11}}\fd G{S_{20}}\right)\right.\nonumber \\[2pt]
&&{} \left.\qquad\qquad {} + 4\,\frac{S_{11}}\rho\left(\fd F{S_{20}}\fd G{S_{02}} - \fd F{S_{02}}\fd G{S_{20}}\right) + 2\,\frac{S_{02}}\rho\left(\fd F{S_{11}}\fd G{S_{02}}-\fd F{S_{02}}\fd G{S_{11}}\right)\right]\nonumber \\[2pt]
&&{}+ \frac{\partial }{\partial z}\left(\frac1\rho\fd F{P_x}\right)\frac1\rho\left(S_{20}\fd G{S_{11}}+2S_{11}\fd G{S_{02}}\right)
-\frac1\rho\left(S_{20}\fd F{S_{11}} + 2S_{11}\fd F{S_{02}}\right)\frac{\partial }{\partial z}\left(\frac1\rho\,\fd G{P_x}\right) \nonumber \\[2pt]
&&{} +\fd F{P_z}\frac1\rho\left(\frac{\partial S_{20}}{\partial z}\fd G{S_{20}} + \frac{\partial S_{11}}{\partial z}\,\fd G{S_{11}} + \frac{\partial S_{02}}{\partial z}\,\fd G{S_{02}}\right)\nonumber \\[2pt]
&&{} \qquad \quad\left.-\frac1\rho\left(\frac{\partial S_{20}}{\partial z}\fd F{S_{20}} + \frac{\partial S_{11}}{\partial z}\fd F{S_{11}} + \frac{\partial S_{02}}{\partial z}\fd F{S_{02}} \right)\fd G{P_z}\right] +\{F,G\}_{\rm c},\label{eqn:Bracket15D}
\end{eqnarray}
where $\{F,G\}_{\rm c}$ is the part of the bracket which explicitly depends on the third order moments $S_{30}$, $S_{21}$, $S_{12}$ and $S_{03}$:
\begin{eqnarray*}
\{F,G\}_{\rm c}&=& \int {\rm d}z\;\frac1\rho \left[S_{30}\left(\frac{\partial}{\partial z}\left(\frac1\rho\fd F{S_{20}}\right)\fd G{S_{11}} - \fd F{S_{11}}\,\frac{\partial}{\partial z}\left(\frac1\rho\fd G{S_{20}}\right)\right)\right.\nonumber \\[2pt]
&&\qquad{}+2S_{21}\left(\frac{\partial}{\partial z}\left(\frac1\rho\,\fd F{S_{20}}\right)\fd G{S_{02}} - \fd F{S_{02}}\frac{\partial}{\partial z}\left(\frac1\rho\fd G{S_{20}}\right)\right)\nonumber\\[2pt]
&&\qquad{} + S_{12}\left(\frac{\partial}{\partial z}\left(\frac1\rho\fd F{S_{02}}\right)\fd G{S_{11}} - \fd F{S_{11}}\,\frac{\partial}{\partial z}\left(\frac1\rho\fd G{S_{02}}\right)\right)\nonumber\\[2pt]
&&\qquad{} + 2S_{03}\left(\frac{\partial}{\partial z}\left(\frac1\rho\fd F{S_{02}}\right)\fd G{S_{02}} - \fd F{S_{02}}\,\frac{\partial}{\partial z}\left(\frac1\rho\fd G{S_{02}}\right)\right)\nonumber\\[2pt]
&&\qquad{} + S_{21}\left(\frac{\partial}{\partial z}\left(\frac1\rho\fd F{S_{11}}\right)\fd G{S_{11}} - \fd F{S_{11}}\frac{\partial}{\partial z}\left(\frac1\rho\fd G{S_{11}}\right)\right)\nonumber\\[2pt]
&&\qquad{}+\left. 2S_{12}\left(\frac{\partial}{\partial z}\left(\frac1\rho\fd F{S_{11}}\right)\fd G{S_{02}} - \fd F{S_{02}}\frac{\partial}{\partial z}\left(\frac1\rho\fd G{S_{11}}\right)\right)\right].
\end{eqnarray*}
As expected, the bracket~(\ref{eqn:Bracket15D}) involving second order moments explicitly contains third order moments as in the one-dimensional case (see Ref.~\onlinecite{Perin14}).  At this stage, it is
natural to consider symmetric distribution functions and assume that these third order moments vanish.  However, it has been noticed in Ref.~\onlinecite{deGuillebon12} that this leads to a failure of the
Jacobi identity and the resulting reduced system is no longer Hamiltonian.  Here we follow a different route, and we consider that the third order moments are closure functions.  In other words, we replace the third
order moments by closure functions of the reduced set of dynamical field variables $\rho$, $P_x$, $P_z$, $S_{20}$, $S_{11}$, $S_{02}$, $E_x$, $E_z$, and $B_y$, i.e.,
\begin{eqnarray*}
&& S_{30}= \Sb 30(\rho, P_x, P_z, S_{20}, S_{11}, S_{02}, E_x, E_z, B_y),\\
&& S_{21}= \Sb 21(\rho, P_x, P_z, S_{20}, S_{11}, S_{02}, E_x, E_z, B_y),\\
&& S_{12}= \Sb 12(\rho, P_x, P_z, S_{20}, S_{11}, S_{02}, E_x, E_z, B_y),\\
&& S_{03}= \Sb 03(\rho, P_x, P_z, S_{20}, S_{11}, S_{02}, E_x, E_z, B_y).
\end{eqnarray*}
The central question is what are appropriate functions $\Sb 30$, $\Sb 21$, $\Sb 12$ and $\Sb 03$, i.e., functions that makes the bracket~(\ref{eqn:Bracket15D}) a Poisson bracket. 

In order to facilitate the calculations, we consider the bracket in terms of the canonical momenta, i.e., we perform the following change of variables
\begin{eqnarray*}
&& \Pi_x=P_x+\frac{q}{c}A_x,\\
&& \Pi_z=P_z+\frac{q}{c}A_z,\\
&& Y=S_{02}\,S_{20}-S_{11}^2,\\
&& \Sigma=\frac{S_{11}}{S_{20}}.
\end{eqnarray*}
The algebra becomes the set of observables $F[\rho,\Pi_x,\Pi_z,S_{20},\Sigma,Y,E_x,E_z,A_x, A_z]$, and the bracket becomes 
\begin{equation}
\begin{split}
\{F,G\}=& \int {\rm d}z \left[\partial_z \fd F\rho \fd G{\Pi_z}-\fd F{\Pi_z}\partial_z \fd G\rho\right.\\[2pt]
& - 4\pi c\left(\fd F{A_x} \fd G{E_x}-\fd F{E_x} \fd G{A_x}+ \fd F{A_z} \fd G{E_z}-\fd F{E_z} \fd G{A_z}\right) \\[2pt]
& + \fd F{\Pi_z}\frac1\rho\left(\partial_z \Pi_x\,\fd G{\Pi_x} + \partial_z S_{20}\,\fd G{S_{20}} + \partial_z \Sigma\,\fd G\Sigma + \partial_z Y\fd G{Y}\right)\\[2pt]
& - \frac1\rho\left(\partial_z \Pi_x\fd F{\Pi_x} + \partial_z S_{20}\fd F{S_{20}} + \partial_z\Sigma\,\fd F\Sigma + \partial_z Y\,\fd F{Y}\right)\fd G{\Pi_z}\\[2pt]
& + \partial_z\left(\frac1\rho\,\fd F{\Pi_x}\right)\frac1\rho\fd G\Sigma-\frac1\rho\,\fd F\Sigma\partial_z\left(\frac1\rho\,\fd G{\Pi_x}\right) - 2\frac{\partial_z \Pi_x}{\rho^2}\left(\fd F{S_{20}}\,G_\Sigma - F_\Sigma\,\fd G{S_{20}}\right) \\[2pt]
& + \frac{\Ub 30}\rho\left(\partial_z \left(\frac1\rho\fd F{S_{20}}\right)\,\fd G\Sigma - \fd F\Sigma\,\partial_z \left(\frac1\rho\,\fd G{S_{20}}\right)\right) \\[2pt]
& + \frac{\Ub 21}\rho\left(\partial_z \left(\frac1\rho\fd F{S_{20}}\right)\,\fd GY - \fd FY\,\partial_z \left(\frac1\rho\fd G{S_{20}}\right) \right) \\[2pt]
& + \frac{\Ub 21}{2\rho\,S_{20}^3} \left(\partial_z \left(\frac1\rho\,\fd F\Sigma\right)\,\fd G\Sigma - \fd F\Sigma\,\partial_z\left(\frac1\rho\,\fd G\Sigma\right) \right)\\[2pt]
& + \frac{\Ub 03}\rho\left[\partial_z \left(\frac1\rho\fd FY\right)\,\fd GY - \fd FY\,\partial_z\left(\frac1\rho\fd GY\right)\right]\\[2pt]
& + \frac1{\rho^2}\left(\frac{Y}{S_{20}}\partial_z \Ub 30+S_{20}^3\partial_z \left(\frac{\Ub 12}{S_{20}^3}\right)\right)\left(\fd F\Sigma\,\fd GY - \fd FY\,\fd G\Sigma\right)\\[2pt]
&\left.  + \frac1\rho\left(\frac{Y}{S_{20}}\,\Ub30 + 3\Ub 12\right)\left[\partial_z \left(\frac1\rho\,\fd F\Sigma\right)\,\fd GY - \fd FY\,\partial_z \left(\frac1\rho\fd G\Sigma\right) \right]\right],
\end{split}\label{eqn:brackfl}
\end{equation}
where the closure functions for this bracket are given by
\begin{eqnarray*}
&& \Ub 30=\frac{\Sb 30}{S_{20}},\\
&& \Ub 21=2S_{20}\left(\Sb 21-\frac{S_{11}}{S_{20}}\Sb 30\right),\\
&& \Ub 12=\Sb 12-2\frac{S_{11}}{S_{20}}\Sb 21+\frac{S_{11}^2}{S_{20}^2}\Sb 30,\\
&& \Ub 03=\frac{Y}{S_{20}}\Ub 21+2S_{20}^2\left( \Sb 03-3\frac{S_{11}}{S_{20}}\Sb 12+3\frac{S_{11}^2}{S_{20}^2}\Sb 21-\frac{S_{11}^3}{S_{20}^3}\Sb 30 \right).
\end{eqnarray*}

\subsection{Constraints from the Jacobi identity}

We require the bracket~(\ref{eqn:brackfl}) to satisfy the Jacobi identity, i.e.,
$$
\{\{F,G\},H\}+\{\{H,F\},G\}+\{\{G,H\},F\}=0,
$$
for all observables $F$, $G$ and $H$, functionals of the field variables $\rho(z)$, $\Pi_x(z)$, $\Pi_z(z)$, $S_{20}(z)$, $\Sigma(z)$, $Y(z)$, $E_x(z)$, $E_z(z)$, $A_x(z)$ and $A_z$. 

By considering $F=\rho(z)$, we deduce that the closure functions do not depend on $\Pi_z$.  By considering $F=\Pi_x(z)$, we deduce that the closure functions do not depend on $\Sigma$.  By considering
$F=\Pi_z(z)$, we deduce that the closure functions do not depend on $\rho$.  By considering $F=E_x(z)$, we deduce that the closure functions do not depend on $A_x$.  By considering $F=A_x(z)$, we
deduce that the closure functions do not depend on $E_x$.  The same holds for $A_z$ and $E_z(z)$.  As a consequence, the closure functions $\Ub 30$, $\Ub 21$, $\Ub 12$
and $\Ub 03$ only depend on $\Pi_x$, $S_{20}$ and $Y$.  Using these results, we identify a sub-algebra of observables $F[\rho,\Pi_x,S_{20},\Sigma,Y]$ with the bracket given by
\begin{equation}
	\begin{split}
\{F,G\} =& \int {\rm d}z\;\frac1\rho \left[ \partial_z\left(\frac{1}{\rho}\,\fd F{\Pi_x}\right)\fd G\Sigma - \fd F\Sigma\,\partial_z\left(\frac1\rho\,\fd G{\Pi_x}\right)\right. \\
& - 2\frac{\partial_z \Pi_x}\rho\left(\fd F{S_{20}}\,\fd G\Sigma - \fd F\Sigma\,\fd G{S_{20}}\right) \\
& + \Ub 30\left(\partial_z \left(\frac1\rho\,\fd F{S_{20}}\right)\fd G\Sigma - \fd F\Sigma\,\partial_z \left(\frac1\rho\,\fd G{S_{20}}\right)\right) \\
&+ \Ub 21\left(\partial_z \left(\frac1\rho\,\fd F{S_{20}}\right)\,\fd GY - \fd FY\,\partial_z \left(\frac1\rho\,\fd G{S_{20}}\right) \right) \\[2pt]
&  +\frac{\Ub 21}{2S_{20}^3} \left(\partial_z \left(\frac1\rho\,\fd F\Sigma\right)\,\fd G\Sigma - \fd F\Sigma\,\partial_z\left(\frac1\rho\,\fd G\Sigma\right)\right)\\
& +\Ub 03\left[\partial_z \left(\frac1\rho\,\fd FY\right)\fd GY - \fd FY\,\partial_z\left(\frac1\rho\,\fd GY\right)\right] \\
&+ \left(\frac{Y}{S_{20}}\,\partial_z\Ub 30 + S_{20}^3\,\partial_z \left(\frac{\Ub 12}{S_{20}^3}\right) \right)\frac1\rho\left(\fd F\Sigma\,\fd GY - \fd FY\,\fd G\Sigma\right) \\
& + \left. \left(\frac{Y}{S_{20}}\,\Ub 30 + 3\Ub 12\right)\left[\partial_z\left(\frac1\rho\,\fd F\Sigma\right)\fd GY - \fd FY\,\partial_z \left(\frac1\rho\,\fd G\Sigma\right)\right]\right].
\end{split}\label{eqn:brackflred}
\end{equation}
A necessary and sufficient condition for the bracket~(\ref{eqn:brackfl}) to be a Poisson bracket is that the bracket~(\ref{eqn:brackflred}) is a Poisson bracket. 

\subsection{A further reduced bracket}
\label{sec:frb}
We consider the subalgebra of observables $F[\rho, \Pi_x,S_{20},Y]$ with the bracket given by
\begin{multline}
\{F,G\}= \int {\rm d}z\;\frac1\rho \left[\Ub 21\left(\partial_z \left(\frac{1}{\rho}\,\fd F{S_{20}}\right)\fd GY - \fd FY\,\partial_z \left(\frac{1}{\rho}\,\fd G{S_{20}}\right)\right)\right.\\
+ \left.\Ub 03\left[\partial_z \left(\frac{1}{\rho}\,\fd FY\right)\,\fd GY - \fd FY\,\partial_z\left(\frac{1}{\rho}\,\fd GY\right)\right]\right]. \label{eqn:redred}
\end{multline}
We notice that this subalgebra only involves the closure functions $\Ub 03$ and $\Ub 21$.  A necessary (but not sufficient) condition for the bracket~(\ref{eqn:brackflred}) to be
a Poisson bracket is that the bracket~(\ref{eqn:redred}) is a Poisson bracket.  A necessary and sufficient condition for the bracket~(\ref{eqn:redred}) to be a Poisson bracket is that the closure
functions $\Ub 21$ and $\Ub 03$ satisfy
\begin{equation}
\label{eqn:JacredYS}
\Ub 21\frac{\partial \Ub 03}{\partial Y}-2\Ub 03\frac{\partial \Ub 21}{\partial Y}-\Ub 21\frac{\partial \Ub 21}{\partial S_{20}}=0.
\end{equation}
We notice that $\Ub 21=0$ is an obvious solution, and in this situation, the bracket~(\ref{eqn:redred}) is a Poisson bracket regardless of the closure function $\Ub 03$.  We
notice that the resulting Poisson bracket has a Casimir invariant, i.e., a function $C$ which commutes with all observables $F$,
$$
\{C,F\}=0,
$$
given by 
$$
C=\int {\rm d}z \rho \int \frac{{\rm d}Y}{\sqrt{\vert \Ub 03 \vert}},
$$
where $\int {\rm d}Y/\sqrt{\vert \Ub 03 \vert}$ is a anti-derivative of $1/\sqrt{\vert \Ub 03 \vert}$ with respect to $Y$.
This invariant defines a normal variable $\phi(\Pi_x,S_{20},Y)=\int {\rm d}Y/\sqrt{\vert \Ub 03 \vert}$. 

If $\Ub 21\not= 0$, we define the functions $\nu(\Pi_x,S_{20},Y)$ and $\eta(\Pi_x,S_{20},Y)$ such that
\begin{eqnarray*}
&& \nu(\Pi_x,S_{20},Y)=\int \frac{{\rm d}Y}{\Ub 21},\\
&& \eta(\Pi_x,S_{20},Y)=\frac{\Ub 03}{\Ub 21^2}.
\end{eqnarray*}
Equation~(\ref{eqn:JacredYS}) becomes
$$
\frac{\partial^2 \nu}{\partial Y\partial S_{20}}+\frac{\partial \eta}{\partial Y}=0,
$$
with solution
$$
\eta(\Pi_x,S_{20},Y)=\Psi(\Pi_x,S_{20})-\frac{\partial \nu}{\partial S_{20}}.
$$ 
This gives the closure
\begin{eqnarray*}
&& \Ub 21=\left(\frac{\partial \nu}{\partial Y} \right)^{-1},\\
&& \Ub 03=\left(\Psi(\Pi_x,S_{20})-\frac{\partial \nu}{\partial S_{20}} \right)\left(\frac{\partial \nu}{\partial Y} \right)^{-2},
\end{eqnarray*} 
where $\Psi$ is an arbitrary function. 
The resulting bracket has a Casimir invariant 
$$
C=\int {\rm d}z \rho \left(\nu-\int {\rm d}S_{20} \Psi\right),
$$
where $\int {\rm d}S_{20} \Psi$ is a anti-derivative of $\Psi$ with respect to $S_{20}$. 
In this case, the closure is more easily expressed in terms of the normal variable $\phi(\Pi_x,S_{20},Y)=\nu-\int {\rm d}S_{20} \Psi$.

In sum, two cases have to be considered, depending on the normal variable $\phi(\Pi_x,S_{20},Y)$: The first case is when 
\begin{eqnarray*}
&& \Ub 21=0,\\
&& \Ub 03=\left(\frac{\partial \phi}{\partial Y} \right)^{-2},
\end{eqnarray*}
and the second case is obtained for
\begin{eqnarray*}
&& \Ub 21=\left(\frac{\partial \phi}{\partial Y} \right)^{-1},\\
&& \Ub 03=-\frac{\partial \phi}{\partial S_{20}} \left(\frac{\partial \phi}{\partial Y} \right)^{-2},
\end{eqnarray*} 
where $\phi$ is an arbitrary function of $\Pi_x$, $S_{20}$ and $Y$.  We notice that these two cases are both constrained up to an arbitrary function $\phi$ of $\Pi_x$, $S_{20}$ and $Y$, and that in
both cases, the corresponding Poisson bracket possesses $\int {\rm d}z \rho \phi$ as a Casimir invariant.

Furthermore, we are looking for a closure which generalizes the one-dimensional case.  From the inspection of the closure functions, we conclude that the first case ($\Ub 21=0$) is the
relevant one since it generalizes the closure found in Ref.~\onlinecite{Perin14}: One feature is that the scaling of the closure function $\Ub 03$ with the normal variable, obtained by replacing
$\phi$ by $u \phi$, is $1/u^2$ as in the one-dimensional case whereas it is $1/u$ in the second case.

\subsection{The case $\Ub 21=0$}

In this case, the bracket~(\ref{eqn:brackflred}) becomes 
\begin{eqnarray*}
\{F,G\}&=& \int {\rm d}z\;\frac1\rho\left[\partial_z\left(\frac1\rho\,\fd F{\Pi_x}\right)\fd G\Sigma - \fd F\Sigma\,\partial_z\left(\frac1\rho\,\fd G{\Pi_x}\right)
- 2\frac{\partial_z\Pi_x}\rho\left(\fd F{S_{20}}\,\fd G\Sigma - \fd F\Sigma\,\fd G{S_{20}}\right) \right. \nonumber \\[2pt]
&&{}  + \Ub 30\left(\partial_z \left(\frac1\rho\,\fd F{S_{20}}\right)\fd G\Sigma - \fd F\Sigma\,\partial_z \left(\frac1\rho\,\fd G{S_{20}}\right)\right)\nonumber\\[2pt]
&&{} + \frac1\rho\left(\frac{Y}{S_{20}}\,\partial_z \Ub 30 + S_{20}^3\partial_z \left(\frac{\Ub 12}{S_{20}^3}\right)\right)\left(\fd F\Sigma\,\fd GY - \fd FY\,\fd G\Sigma\right)\nonumber\\[2pt]
&&{}   + \left(\frac{Y}{S_{20}} \Ub 30+3\Ub 12 \right) \left[\partial_z \left(\frac1\rho\,\fd F\Sigma\right)\fd GY - \fd FY\,\partial_z\left(\frac1\rho\,\fd G\Sigma\right) \right]\nonumber\\[2pt]
&&{} + \left. \Ub 03\left[\partial_z \left(\frac1\rho\,\fd FY\right)\fd GY - \fd FY\,\partial_z\left(\frac1\rho\,\fd GY\right)\right]\right]. 
\end{eqnarray*}
Inserting the above-bracket into the Jacobi identity leads to a set of coupled nonlinear partial differential equations in the closure functions.  The number of these equations depends on the number
of field variables.  Here there are four field variables, so we expect a dozen of coupled PDEs, among which some are not necessarily independent from the others.  It is too lengthy to list all of these conditions, and not necessary for the approach we have chosen:
We first consider a subset of these conditions, find solutions of this subset of equations, form an ansatz for the closure functions, and then insert it back into the Jacobi identity to determine further constraints.  We iterate this procedure until the bracket satisfies the Jacobi identity.  Here we begin with the following set of conditions required by the Jacobi
identity:
\begin{eqnarray}
&& \Ub 03\frac{\partial \Ub 30}{\partial Y}=0,\label{eqn:U03}\\
&& \Ub 30\left(\frac{Y}{S_{20}}\frac{\partial \Ub 30}{\partial Y}+\frac{\partial \Ub 30}{\partial S_{20}} \right)+\frac{\partial \Ub 30}{\partial \Pi_x}+3\Ub 12\frac{\partial \Ub 30}{\partial Y}+2=0,\label{eqn:U30}\\
&& \Ub 30\left(\frac{Y}{S_{20}}\frac{\partial \Ub 12}{\partial Y}+\frac{\partial \Ub 12}{\partial S_{20}} \right)+\frac{\partial \Ub 12}{\partial \Pi_x}+3\Ub 12\frac{\partial \Ub 12}{\partial Y}-3\frac{\Ub 12 \Ub 30}{S_{20}}-2\frac{Y}{S_{20}}=0,\label{eqn:U12}\\
&& \Ub 30\left(\frac{Y}{S_{20}}\frac{\partial \Ub 03}{\partial Y}+\frac{\partial \Ub 03}{\partial S_{20}} \right)+\frac{\partial \Ub 03}{\partial \Pi_x}+3\Ub 12\frac{\partial \Ub 03}{\partial Y}-2\frac{\Ub 03 \Ub 30}{S_{20}}-4\Ub 03\frac{\partial \Ub 12}{\partial Y}=0.\label{eqn:U03_fin}
\end{eqnarray}
From Eq.~(\ref{eqn:U03}), we deduce that $\Ub 03=0$ or $\Ub 30$ is independent of $Y$. 
We investigate a closure for which $\Ub 03\not= 0$ in analogy with the one-dimensional case for which the third order moment is an arbitrary function of the second order moment.\cite{Perin14} Equation~(\ref{eqn:U30}) reduces to 
\begin{equation}
\label{eqn:Burgers}
\Ub 30 \frac{\partial \Ub 30}{\partial S_{20}} +\frac{\partial \Ub 30}{\partial \Pi_x}+2=0,
\end{equation}
which resembles an inviscid Burgers' equation. In particular, there are two obvious solutions, one dependent on $\Pi_x$ only and the other one dependent on $S_{20}$ only:
\begin{eqnarray*}
&& \Ub 30^{(1)}=-2\Pi_x+2\alpha,\\
&& \Ub 30^{(2)}=\sqrt{\kappa -4S_{20}},
\end{eqnarray*}   
but we will see that none of these obvious solutions lead to a Hamiltonian closure. 

Since $\Ub 30$ is independent of $Y$, an additional constraint resulting from the Jacobi identity is given by
$$
S_{20}\left(\frac{\partial \Ub 12}{\partial Y} \right)^2 -\Ub 30\frac{\partial \Ub 12}{\partial Y}-1=0.
$$
From this equation, we conclude that $\partial \Ub 12/\partial Y$ is independent of $Y$, i.e., 
$$
\Ub 12=Y\overline{V}_{12}(\Pi_x,S_{20})+\overline{W}_{12}(\Pi_x,S_{20}),
$$ 
where $\overline{V}_{12}$ satisfies $S_{20}\overline{V}_{12}^2-\Ub 30\overline{V}_{12}-1=0$. We insert this equation for $\Ub 12$ into Eq.~(\ref{eqn:U12}) and we obtain the following two equations
\begin{eqnarray*}
&& \Ub 30\frac{\partial \overline{V}_{12}}{\partial S_{20}}+\frac{\partial \overline{V}_{12}}{\partial \Pi_x}+\overline{V}_{12}^2=0,\\
&& \Ub 30\frac{\partial \overline{W}_{12}}{\partial S_{20}}+\frac{\partial \overline{W}_{12}}{\partial \Pi_x}+3\overline{W}_{12}\left(\overline{V}_{12}-\frac{\Ub 30}{S_{20}} \right)=0.
\end{eqnarray*}
The first equation is always satisfied provided that $S_{20}\overline{V}_{12}^2-U_{30}\overline{V}_{12}-1=0$ and $\Ub 30$ satisfies Eq.~(\ref{eqn:U30}).  We solve these equations by assuming
that the dependence on $\Pi_x$ of the functions $\overline{V}_{12}$ and $\overline{W}_{12}$ solely comes from the dependence of $\Ub 30$ on $\Pi_x$, i.e., we assume that $\overline{V}_{12}$
and $\overline{W}_{12}$ are functions of $S_{20}$ and $\Ub 30$.  In this way, we have
\begin{eqnarray}
&& \overline{V}_{12}=\frac{\Ub 30\pm\sqrt{\Ub 30^2+4S_{20}}}{2S_{20}},\label{eqn:V12} \\
&& \overline{W}_{12}=\left(\Ub 30\pm \sqrt{\Ub 30^2+4S_{20}}\right)^3\Phi_{12}(\Ub 30^2+4S_{20}),\label{eqn:W12}
\end{eqnarray}
where $\Phi_{12}$ is an arbitrary scalar function. We insert these expressions in the Jacobi identity. The constraint on $\Ub 12$ imposes that $\Phi_{12}=0$, and an additional constraint on $\Ub 30$ is given by
$$
2S_{20}\frac{\partial \Ub 30}{\partial S_{20}}+\sqrt{\Ub 30^2+4S_{20}}-\Ub 30=0.
$$
The solution of this equation is given by
$$
\Ub 30=\frac{1}{\beta(\Pi_x)}-\beta(\Pi_x)S_{20},
$$
i.e., the closure function $\Ub 30$ is linear in $S_{20}$. We insert this equation into Eq.~(\ref{eqn:Burgers}) and we find
$$
\Ub 30=\frac{\left(\Pi_x-\alpha \right)^2-S_{20}}{\alpha-\Pi_x},
$$
where $\alpha$ is an arbitrary constant. Considering the cases $\Pi_x<\alpha$ and $\Pi_x>\alpha$ in Eqs.~(\ref{eqn:V12}-\ref{eqn:W12}), we show that the two branches for $\Ub 12$ lead to the same solution
$$
\Ub 12=Y\frac{\alpha-\Pi_x}{S_{20}}.
$$
Next, we solve Eq.~(\ref{eqn:U03_fin}) for $\Ub 03$. Since this equation is a first order linear partial differential equation, its solution is obtained using the method of characteristics, and it is given by
\begin{equation}
\label{eqn:formU03}
\Ub 03=\frac{S_{20}^6}{\left(\Pi_x-\alpha \right)^4}\Phi_{03}\left(\frac{\left(\Pi_x-\alpha \right)^2+S_{20}}{\alpha-\Pi_x},\frac{Y}{S_{20}^4}\left(\Pi_x-\alpha \right)^3\right),
\end{equation}
where $\Phi_{03}$ is an arbitrary function of two variables. An additional condition resulting from the Jacobi identity implies that
$$
x\frac{\partial \Phi_{03}}{\partial x}-3y\frac{\partial \Phi_{03}}{\partial y}+4\Phi_{03}=0,
$$
leading to
$$
\Ub 03=2S_{20}^2 \left(\frac{Y}{S_{20}} \right)^{4/3}\Phi\left(\left( \frac{Y}{S_{20}}\right)^{1/3}\frac{ \left(\Pi_x-\alpha\right)^2+S_{20}}{S_{20}} \right),
$$
where $\Phi$ is an arbitrary function. 

In summary, the closure for the bracket~(\ref{eqn:Bracket15D}) is given by
\begin{eqnarray}
&&  \Sb 30=S_{20}(\alpha-\Pi_x)-\frac{S_{20}^2}{\alpha -\Pi_x},\label{eqn:S30}\\
&&  \Sb 21=S_{11}(\alpha-\Pi_x)-\frac{S_{20} S_{11}}{\alpha -\Pi_x}, \label{eqn:S21}\\
&&  \Sb 12=S_{02}\left(\alpha-\Pi_x\right)-\frac{S_{11}^2}{\alpha-\Pi_x},\label{eqn:S12}\\
&&  \Sb 03=\frac{S_{11}}{S_{20}}\left(3S_{02}-2\frac{S_{11}^2}{S_{20}} \right)(\alpha - \Pi_x)-\frac{S_{11}^3}{S_{20}(\alpha -\Pi_x)}\nonumber\\
&&    \qquad +\left(S_{02}-\frac{S_{11}^2}{S_{20}} \right)^{4/3} \Phi\left(\left( S_{02}-\frac{S_{11}^2}{S_{20}}\right)^{1/3}\frac{ \left( \Pi_x-\alpha\right)^2+S_{20} }{S_{20}} \right),\label{eqn:S03}
\end{eqnarray}
where $\Phi$ is an arbitrary function. In these expressions, $\Pi_x=P_x+(q/c)\int {\rm d}z B_y$ is the canonical momentum. When we insert this closure into the Jacobi identity, we show that this identity vanishes. In order to check this, the Mathematica~\cite{Mathematica} notebook is provided at \texttt{github.com/cchandre/VM15D}.  

{\em Remark: Comparison with the one-dimensional case --} 
For $f(z,p_x,p_z)=f_{1D}(z,p_z)\delta(p_x)$, we have $P_x=0$, $S_{11}=0$ and $S_{20}=0$. The closure is on the moment $\Sb 03$ and it is given by
$$
\Sb 03=\Phi(S_{02}).
$$
Therefore we recover the one-dimensional closure which states that the moment of order three is given by an arbitrary function of the moment of order two. 

Next, we determine the Casimir invariants of the Poisson bracket that are of the entropy-form:
$$
C=\int {\rm d}z\ \rho\ \phi(\Pi_x,S_{20},\Sigma,Y).
$$
From $\{\Pi_x,C\}=0$ and $\{S_{20},C\}=0$, we conclude that $\phi$ does not depend on $\Sigma$. From $\{Y,C\}=0$, we obtain
$$
\Ub 03\left(\frac{\partial \phi}{\partial Y} \right)^2=r,
$$
where $r=0$ or $r=\pm 1$ (depending on the sign of $\Ub 03$).  We first consider the case $r=0$.  The Casimir invariant must satisfy $\{\Sigma,C\}=0$ which reduces to
$$
\partial_z \frac{\partial \phi}{\partial \Pi_x}-2\partial_z \Pi_x\frac{\partial \phi}{\partial S_{20}}+\Ub 30\partial_z \frac{\partial \phi}{\partial S_{20}}=0.
$$ 
Since $\Ub 30$ does not depend on $Y$, this equation has a solution given by
$$
\phi=\left((\Pi_x-\alpha)^2+S_{20} \right)\Gamma \left(\frac{\Pi_x-\alpha}{(\Pi_x-\alpha)^2+S_{20}} \right),
$$
where $\Gamma$ is an arbitrary function.  This constitutes an infinite family of Casimir invariants.  They are obviously absent in the one dimensional case, since they only depend on $\Pi_x$ and
$S_{20}$.  In particular, the total momentum, obtained for $\Gamma(p)=p$, is a Casimir invariant corresponding to translation invariance in $x$.

The case $r=\pm 1$ leads to 
$$
\phi=\int \frac{{\rm d Y}}{\sqrt{\vert \Ub 03\vert}},
$$
as in the first case of Sec.~\ref{sec:frb}.
This Casimir invariant is a generalization of the Casimir invariant found in the one-dimensional system. It represents the total entropy in the longitudinal direction.\cite{Perin14} The relation between $\phi$ and the closure function $\Ub 03$ given by Eq.~(\ref{eqn:formU03}) is
$$
\phi=\frac{S_{20}}{(\Pi_x-\alpha)^2+S_{20}}\Psi\left(\left( \frac{Y}{S_{20}}\right)^{1/3}\frac{ \left( \Pi_x-\alpha\right)^2+S_{20} }{S_{20}}  \right),
$$
where $(\Psi^\prime)^2\Phi={\rm const}$.

In summary, the following observables are Casimir invariants of the bracket~(\ref{eqn:Bracket15D}):
\begin{eqnarray}
&& C_{\rm t}=\int {\rm d}z \rho \left((\Pi_x-\alpha)^2+S_{20} \right)\Gamma \left(\frac{\Pi_x-\alpha}{(\Pi_x-\alpha)^2+S_{20}} \right),\label{eqn:Ct}\\
&& C_{\rm l}=\int {\rm d}z \rho \frac{S_{20}}{(\Pi_x-\alpha)^2+S_{20}}\Psi\left(\left( S_{02}-\frac{S_{11}^2}{S_{20}}\right)^{1/3}\frac{ \left( \Pi_x-\alpha\right)^2+S_{20} }{S_{20}}   \right),\label{eqn:Cl}
\end{eqnarray}
where $\Gamma$ is an arbitrary function and $\Psi$ is determined by the choice of the closure function $\Sb 03$, i.e., $(\Psi^\prime)^2\Phi={\rm const}$. 

{\em Remark: Gauge invariance -- } Since the closure functions depend on $\Pi_x=P_x+(q/c) A_x$, it might be concluded that the closure depends on the gauge for the vector potential, i.e., on the
change $A_x\mapsto A_x+a_0 $.  However, by adjusting the free parameter $\alpha$, this gauge can be fixed and the closure functions made gauge free.

If we require that the longitudinal entropy is not a function of the fields $B_y$ and the transverse momentum $P_x$, $\Psi$ has to be chosen as $\Psi(p)=p$ (which implies that $\Phi$ is a constant). The longitudinal entropy is then given by
$$
C_{\rm l}=\int {\rm d}z \rho \left(S_{02}-\frac{S_{11}^2}{S_{20}} \right)^{1/3}. 
$$
The closure function $\Sb 03$ becomes
$$
\Sb 03=\frac{S_{11}}{S_{20}}\left(3S_{02}-2\frac{S_{11}^2}{S_{20}} \right)(\alpha - \Pi_x)-\frac{S_{11}^3}{S_{20}(\alpha -\Pi_x)}+\lambda \left(S_{02}-\frac{S_{11}^2}{S_{20}} \right)^{4/3},
$$
where $\lambda$ is an arbitrary constant. This closure corresponds to a more physical model in which the longitudinal entropy density does not depend on the momentum in the transverse direction. In other words, we have used the transverse direction to specify the closure in the longitudinal direction.

\subsection{The case $\Ub 21\not= 0$}

The closure functions need to satisfy the following constraints:
\begin{eqnarray}
&& \Ub 21\left(\frac{Y}{S_{20}}\frac{\partial \Ub 30}{\partial Y}-\frac{\partial \Ub 30}{\partial S_{20}}+\frac{\partial \Ub 12}{\partial Y} \right) -2\Ub 03\frac{\partial \Ub 30}{\partial Y}=0,\label{eqn:U03_nz}\\
&& \Ub 30\left(\frac{Y}{S_{20}}\frac{\partial \Ub 30}{\partial Y}+\frac{\partial \Ub 30}{\partial S_{20}} \right)+\frac{\partial \Ub 30}{\partial \Pi_x}+3\Ub 12\frac{\partial \Ub 30}{\partial Y}-\frac{\Ub 21}{2S_{20}^3}\frac{\partial \Ub 21}{\partial Y}+2=0,\label{eqn:U30_nz}\\
&& \Ub 21\left(\frac{\partial \Ub 21}{\partial S_{20}}-\frac{\partial \Ub 03}{\partial Y} \right)+2\Ub 03\frac{\partial \Ub 21}{\partial Y}=0,\\
&& \Ub 30\left(\frac{Y}{S_{20}}\frac{\partial \Ub 12}{\partial Y}+\frac{\partial \Ub 12}{\partial S_{20}} \right)+\frac{\partial \Ub 12}{\partial \Pi_x}+3\Ub 12\frac{\partial \Ub 12}{\partial Y}-3\frac{\Ub 12\ \Ub 30}{S_{20}}-2\frac{Y}{S_{20}}\nonumber \\
&& \qquad +\frac{\Ub 21}{2S_{20}^3}\left(\frac{Y}{S_{20}}\frac{\partial \Ub 21}{\partial Y}-\frac{\partial \Ub 03}{\partial Y}+\frac{3\Ub 21}{S_{20}} \right)=0,\label{eqn:U12_nz}\\
&& \Ub 30\left(\frac{Y}{S_{20}}\frac{\partial \Ub 03}{\partial Y}+\frac{\partial \Ub 03}{\partial S_{20}} \right)+\frac{\partial \Ub 03}{\partial \Pi_x}+3\Ub 12\frac{\partial \Ub 03}{\partial Y}-2\frac{\Ub 03\ \Ub 30}{S_{20}}-4\Ub 03\frac{\partial \Ub 12}{\partial Y}\nonumber \\
&& \qquad +\frac{\Ub 21}{S_{20}}\left(\frac{Y}{S_{20}}\Ub 30-2S_{20}\frac{\partial \Ub 12}{\partial S_{20}}-3\Ub 12 \right)=0,\\
&& \Ub 30\left(\frac{Y}{S_{20}}\frac{\partial \Ub 21}{\partial Y}+\frac{\partial \Ub 21}{\partial S_{20}} \right)+\frac{\partial \Ub 21}{\partial \Pi_x}+3\Ub 12\frac{\partial \Ub 21}{\partial Y}-\frac{\Ub 21\ \Ub 30}{S_{20}}-2\Ub 21\frac{\partial \Ub 12}{\partial Y}=0,\label{eqn:U03_nz_fin}
\end{eqnarray}
which are generalizations of Eqs.~(\ref{eqn:U03}-\ref{eqn:U03_fin}) to the case $\Ub 12\not= 0$. 
These equations are more cumbersome to solve since there are more nonlinearities than in Eqs.~(\ref{eqn:U03}-\ref{eqn:U03_fin}). Here we are looking for solutions which lead to a systems possessing entropy-like Casimir invariants, i.e., of the form
$$
C=\int {\rm d}z \rho \phi(\Pi_x,S_{20},Y).
$$
This leads to three additional constraints on the closure functions obtained from $\{S_{20},C\}=0$, $\{Y,C\}=0$ and $\{\Sigma,C\}=0$:
\begin{eqnarray}
&& \Ub 21\frac{\partial \phi}{\partial Y}={\rm const},\\
&& \Ub 03\left(\frac{\partial \phi}{\partial Y} \right)^2+\frac{\partial \phi}{\partial S_{20}}={\rm const},\\
&& \Ub 30 \partial_z \left(\frac{Y}{S_{20}}\frac{\partial \phi }{\partial Y}+\frac{\partial \phi}{\partial S_{20}} \right)+\partial_z \frac{\partial \phi}{\partial \Pi_x}-2\partial_z \Pi_x \frac{\partial \phi}{\partial S_{20}}+3\partial_z \left(U_{12}\frac{\partial \phi}{\partial Y} \right)-S_{20}^3\frac{\partial \phi}{\partial Y}\partial_z \frac{\Ub 12}{S_{20}^3}=0. \label{eqn:CasU21}
\end{eqnarray}
The first two equations determine $\Ub 03$ and $\Ub 21$ as functions of $\phi$, while the last equations only involves the closure functions $\Ub 30$ and $\Ub 12$.  We notice that this last equation
is exactly the same as in the case $\Ub 21=0$.  For this equation, we found that a physically relevant closure with $\phi=\tilde{\phi}(Y/S_{20})$.  This corresponds to having an entropy-like Casimir
invariant which is independent of $\Pi_x$ and it corresponds to the Casimir invariant found in the one-dimensional case.  The question is then: Are there other closure functions $\Ub 30$ and $\Ub 12$
for which $\phi(\Pi_x,S_{20},Y)=\tilde{\phi}(Y/S_{20})$ is a solution of Eq.~(\ref{eqn:CasU21})?  This equation becomes
$$
3\partial_z\left(\tilde{\phi}'\frac{\Ub 12}{S_{20}} \right)-S_{20}^2\tilde{\phi}'\partial_z \frac{\Ub 12}{S_{20}^3}+\frac{2Y}{S_{20}^2}\tilde{\phi}'\partial_z \Pi_x=0.
$$
By looking at the terms in $\partial_z \Pi_x$, we have
$$
\frac{\partial \Ub 12}{\partial \Pi_x}=-\frac{Y}{S_{20}}, 
$$
and by looking at the terms in $\partial_z S_{20}$ and $\partial_z Y$, we conclude that the function $({\widetilde{\phi}}')^{3/2}\Ub 12$ is independent of $S_{20}$ and $Y$. By combining these two observations, we conclude that the only possible solution is
$$
\Ub 12=\frac{Y}{S_{20}}(\alpha-\Pi_x),
$$
together with 
$$
\widetilde{\phi}(x)=3\kappa x^{1/3},
$$
as in the case $\Ub 21=0$. We notice that this provides an expression for $\Ub 21$ and $\Ub 03$ given by
\begin{eqnarray*}
&& \Ub 21= \kappa^{-1}S_{20}^{1/3}Y^{2/3},\\
&& \Ub 03= \Psi S_{20}^{2/3}Y^{4/3}+\kappa^{-1}S_{20}^{-2/3}Y^{5/3}.
\end{eqnarray*}
Inserting this closure in Eq.~(\ref{eqn:U12_nz}) gives an expression for $\Ub 30$:
$$
\Ub 30=-\frac{(\Pi_x-\alpha)^2-S_{20}}{\Pi_x-\alpha}-\frac{Y^{1/3}S_{20}^{-4/3}}{3\kappa^2(\Pi_x-\alpha)}+\frac{2\Psi}{9\kappa(\Pi_x-\alpha)}. 
$$ 
These expressions do not satisfy Eq.~(\ref{eqn:U03_nz}). We can check that this closure does not satisfy the Jacobi identity. 

Of course, we cannot rule out possible Hamiltonian closures which do not have any Casimir invariants of the form $\int {\rm d}z \rho \tilde{\phi}(Y/S_{20})$, but we argue that these closures are less relevant.  In any case, as stated above, the closures with $\Ub 21\not= 0$ would not correspond to generalizations of the purely longitudinal case.

\section{Hamiltonian warm-fluid model: Poisson bracket, Hamiltonian, equations of motion and linearization around homogeneous equilibria}
\label{sec:eqn}

In summary, the only possible Hamiltonian models resulting from the closure of the third order moments are given by the following Poisson bracket
\begin{eqnarray*}
\{F,G\}&=& \int {\rm d}z \left[\left( \frac{\partial}{\partial z} \fd F\rho - 4\pi q \fd F{E_z}\right)\fd G{P_z}-\fd F{P_z}\left( \frac{\partial }{\partial z}\fd G\rho - 4\pi q \fd G{E_z}\right)\right.  \\[2pt]
&& +4\pi\left(  c\frac{\partial }{\partial z}\fd F{B_y} + q \fd F{P_x}\right)\fd G{E_x}-4\pi \fd F{E_x}\left( c\frac{\partial}{\partial z}\,\fd G{B_y} + q \fd G{P_x}\right)  \\[2pt]
&& -\left(\frac{q B_y}{c}+\frac{\partial P_x}{\partial z} \right)\frac1\rho\left[\fd F{P_x}\,\fd G{P_z} - \fd F{P_z}\,\fd G{P_x} + 2\,\frac{S_{20}}\rho\left(\fd F{S_{20}}\,\fd G{S_{11}} - \fd F{S_{11}}\,\fd G{S_{20}}\right)\right.  \\[2pt]
&& \left. \qquad \qquad +4\,\frac{S_{11}}\rho\left(\fd F{S_{20}}\,\fd G{S_{02}} - \fd F{S_{02}}\,\fd G{S_{20}}\right) + 2\,\frac{S_{02}}\rho\left(\fd F{S_{11}}\,\fd G{S_{02}} - \fd F{S_{02}}\,\fd G{S_{11}}\right)\right]  \\[2pt]
&&+ \frac{\partial }{\partial z}\left(\frac1\rho\,\fd F{P_x}\right)\left(S_{20}\fd G{S_{11}} + 2 S_{11}\fd G{S_{02}}\right)
- \frac1\rho\left(S_{20}\fd F{S_{11}} + 2 S_{11} \fd F{S_{02}}\right)\frac{\partial }{\partial z}\left(\frac1\rho\,\fd G{P_x}\right)   \\[2pt]
&& +\fd F{P_z}\frac1\rho\left(\frac{\partial S_{20}}{\partial z}\,\fd G{S_{20}} + \frac{\partial S_{11}}{\partial z}\,\fd G{S_{11}} + \frac{\partial S_{02}}{\partial z}\fd G{S_{02}}\right)  \\[2pt]
&& \qquad \quad\left.-\frac1\rho\left(\frac{\partial S_{20}}{\partial z}\fd F{S_{20}} + \frac{\partial S_{11}}{\partial z} \fd F{S_{11}} + \frac{\partial S_{02}}{\partial z}\fd F{S_{02}}\right)\fd G{P_z}\right] +\{F,G\}_{\rm c}\,, 
\end{eqnarray*}
where $\{F,G\}_{\rm c}$ is given by
\begin{align*}
\{F,G\}_{\rm c}=& \int{\rm d}z\;\frac1\rho \left[ \Sb 30\left(\frac{\partial}{\partial z}\left(\frac1\rho\fd F{S_{20}}\right)\fd G{S_{11}} - \fd F{S_{11}}\,\frac{\partial}{\partial z}\left(\frac1\rho\fd G{S_{20}}\right)\right)\right.  \\[2pt]
&+2\Sb 21\left(\frac{\partial}{\partial z}\left(\frac1\rho\,\fd F{S_{20}}\right)\fd G{S_{02}} - \fd F{S_{02}}\,\frac{\partial}{\partial z}\left(\frac1\rho\,\fd G{S_{20}}\right)\right) \\[2pt]
& + \Sb 12\left(\frac{\partial}{\partial z}\left(\frac1\rho\,\fd F{S_{02}}\right)\fd G{S_{11}} - \fd F{S_{11}}\,\frac{\partial}{\partial z}\left(\frac1\rho\,\fd G{S_{02}}\right)\right) \\[2pt]
& + 2\Sb 03\left(\frac{\partial}{\partial z}\left(\frac1\rho\,\fd F{S_{02}}\right)\fd G{S_{02}} - \fd F{S_{02}}\,\frac{\partial}{\partial z}\left(\frac1\rho\,\fd G{S_{02}}\right)\right) \\[2pt]
&  +\Sb 21\left(\frac{\partial}{\partial z}\left(\frac1\rho\,\fd F{S_{11}}\right)\fd G{S_{11}} - \fd F{S_{11}}\,\frac{\partial}{\partial z}\left(\frac1\rho\,\fd G{S_{11}}\right)\right) \\[2pt]
&\left.+2\Sb 12\left(\frac{\partial}{\partial z}\left(\frac1\rho\,\fd F{S_{11}}\right)\fd G{S_{02}} - \fd F{S_{02}}\,\frac{\partial}{\partial z}\left(\frac1\rho\,\fd G{S_{11}}\right)\right)\right].
\end{align*}
Here the closure functions are given by 
\begin{eqnarray*}
&&  \Sb 30=S_{20}(\alpha-\Pi_x)-\frac{S_{20}^2}{\alpha -\Pi_x}, \\
&&  \Sb 21=S_{11}(\alpha-\Pi_x)-\frac{S_{20} S_{11}}{\alpha -\Pi_x},  \\
&&  \Sb 12=S_{02}\left(\alpha-\Pi_x\right)-\frac{S_{11}^2}{\alpha-\Pi_x},\\
&&  \Sb 03=\frac{S_{11}}{S_{20}}\left(3S_{02}-2\frac{S_{11}^2}{S_{20}} \right)(\alpha - \Pi_x)-\frac{S_{11}^3}{S_{20}(\alpha -\Pi_x)} \\
&&    \qquad +\left(S_{02}-\frac{S_{11}^2}{S_{20}} \right)^{4/3} \Phi\left(\left( S_{02}-\frac{S_{11}^2}{S_{20}}\right)^{1/3}\frac{ \left( \Pi_x-\alpha\right)^2+S_{20} }{S_{20}} \right), 
\end{eqnarray*}
where $\Phi$ is an arbitrary function and $\alpha$ an arbitrary constant. In the expressions of the closure functions, $\Pi_x=P_x+(q/c)\int {\rm d}z B_y$ is the canonical momentum. 

In the non-relativistic case, the kinetic Hamiltonian, given by
$$
H=\int {\rm d}z {\rm d}p_x {\rm d}p_z\;f\,\frac{p_x^2+p_z^2}{2m} + \int{\rm d}z\;\frac{E_x^2+E_z^2+B_y^2}{8\pi},
$$
becomes 
\begin{equation}
H=\int {\rm d}z\left[\rho\,\frac{P_x^2+P_z^2}{2m}+\frac{\rho}{2m}\left(S_{20}+\rho^2 S_{02}\right) + \frac{E_x^2+E_z^2+B_y^2}{8\pi} \right],\label{eqn:Ham}
\end{equation}
without any approximations.  In other words, there is no reduction of the Hamiltonian, given that $H$ only depends on the variables $\rho$, $P_x$, $P_z$, $S_{20}$, $S_{11}$, $S_{02}$, and the fields
$E_x$, $E_z$ and $B_y$.  Hamiltonian~(\ref{eqn:Ham}) with the bracket~(\ref{eqn:Bracket15D}) lead to the following equations of motion:
\begin{eqnarray}
&& \dot{\rho}=-\partial_z \left(\rho \frac{P_z}{m} \right),\label{eqn:mot_d}\\
&& \dot{P}_x=-\frac{P_z}{m}\partial_z P_x +qE_x -\frac{qP_z B_y}{mc}-\frac1\rho\partial_z \left(\frac{\rho^2 S_{11}}{m} \right),\\
&& \dot{P}_z=-\frac{P_z}{m}\partial_z P_z +qE_z +\frac{qP_x B_y}{mc}-\frac1\rho\partial_z \left(\frac{\rho^3 S_{02}}{m} \right),\\
&& \dot{S}_{20} = -\frac{P_z}{m}\partial_z S_{20}-\frac{2\rho S_{11}}{m}\left(\frac{qB_y}{c}+\partial_z P_x \right)-\frac1\rho\partial_z \left( \frac{\rho^2 \Sb 21}{m}\right),\\
&& \dot{S}_{11} = -\frac{P_z}{m}\partial_z S_{11}+\frac{q B_y S_{20}}{\rho m c}-\frac{\rho S_{02}}{m}\left(\frac{qB_y}{c}+\partial_z P_x \right)-\frac{1}{\rho^2}\partial_z \left( \frac{\rho^3 \Sb 12}{m}\right), \\
&& \dot{S}_{02} = -\frac{P_z}{m}\partial_z S_{02}+\frac{2qB_yS_{11}}{\rho m c}-\frac{1}{\rho^3}\partial_z\left(\frac{\rho^4\Sb 03}{m} \right),\\
&& \dot{E}_x =-c\partial_z B_y-\frac{4\pi q \rho P_x}{m},\\
&& \dot{E}_z = -\frac{4\pi q \rho P_z}{m},\\
&& \dot{B}_y =-c\partial_z E_x.\label{eqn:mot_f}
\end{eqnarray}
These equations depend on the choice of one constant $\alpha$ and one scalar function of one variable $\Phi$. The equations of motion~(\ref{eqn:mot_d}-\ref{eqn:mot_f}) possess conserved quantities given by Eqs.~(\ref{eqn:Ct}) and (\ref{eqn:Cl}).

A simpler model is obtained by considering that $\Phi$ is a constant, i.e.,
$\Phi(x)=\lambda$ (and hence there exist a Casimir invariant of the entropy type which is independent of the momenta), leading to a Hamiltonian warm fluid model defined by two arbitrary constants,
$\alpha$ and $\lambda$.  

\begin{figure}[hbt]
\centering
\includegraphics[width=8cm]{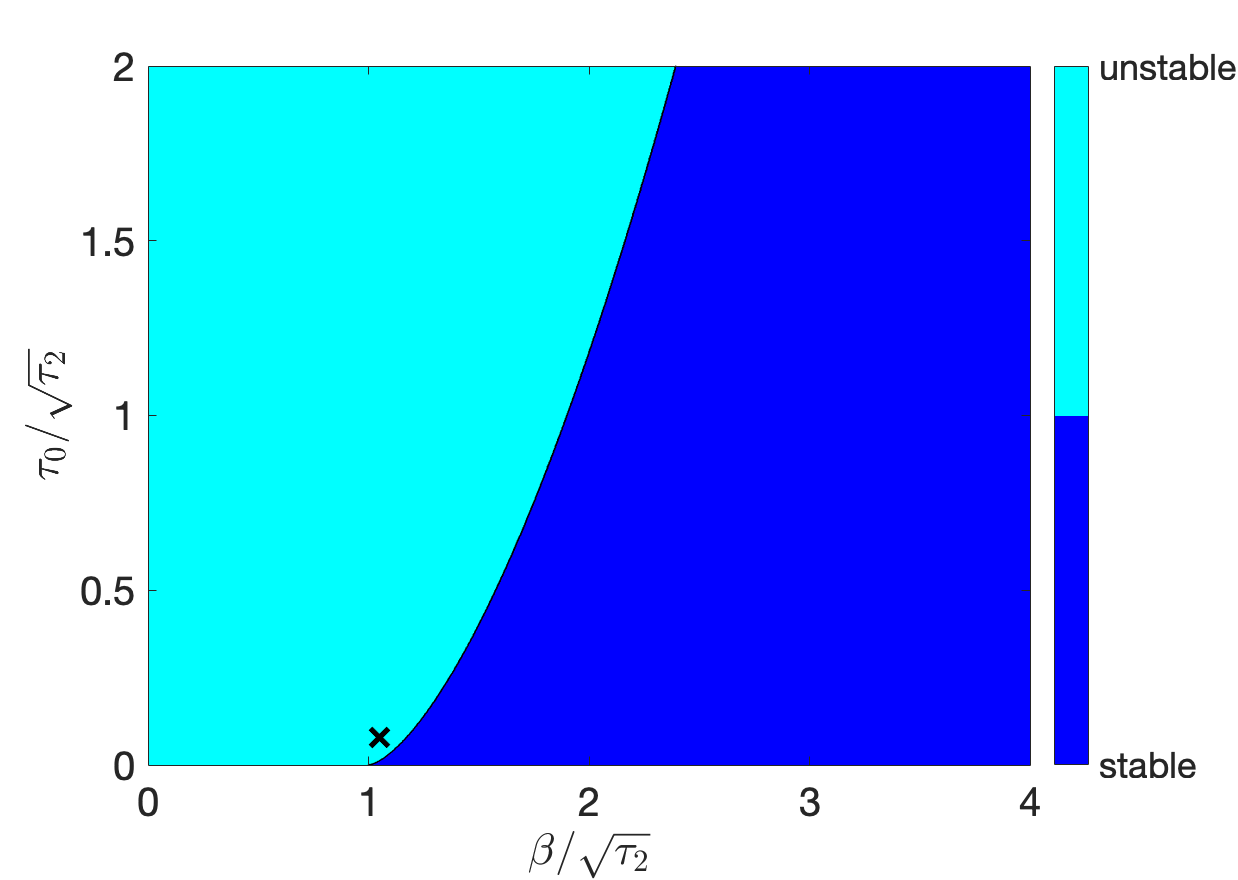}
\caption{\label{fig:smallk} Set of parameters $(\beta,\tau_0)$ for which the small-$k$ modes are unstable. The black cross indicates the chosen parameters for the solutions of the dispersion relation~\eqref{dispersion} depicted in Fig.~\ref{fig:dispersion}.}
\end{figure}

Depending on the values of the parameters $\alpha$ and $\lambda$, the resulting model might exhibit very different dynamics, even qualitatively.  The choice of these parameters are guided by some of
the features we would like to have reproduced from the kinetic model.  This is the advantage of having parameters in the closure.  Here we look at the linearization of these Hamiltonian reduced models
near homogeneous equilibria.  More precisely, we consider the transverse momenta and fields and investigate the conditions under which these homogeneous equilibria are unstable for small wavelengths 
\cite{Weibel:1959aa}:
\begin{eqnarray*}
&&\rho(z,t)=\rho_0,\\
&& P_x(z,t)= \delta P_x {\rm e}^{i (k z-\omega t)},\\
&& P_z(z,t)=0,\\
&& S_{20}(z,t)= m\,T_{20}+\delta S_{20}{\rm e}^{i (k z-\omega t)},\\
&& S_{11}(z,t)= \delta S_{11}{\rm e}^{i k z},\\
&& S_{02}(z,t)=\rho_0^{-2\,}m\,T_{02} +\delta S_{02} {\rm e}^{i (k z-\omega t)},\\
&& E_x(z,t)= \delta E_x {\rm e}^{i (k z-\omega t)},\\
&& E_z(z,t)=0,\\
&& B_y(z,t)=\delta B_y{\rm e}^{i (k z-\omega t)}.
\end{eqnarray*}   
The dispersion relation $\omega(k)=\omega_p\,X(k)$ (where $\omega_p=\sqrt{4\pi \rho_0 q^2/m}$ is the plasma frequency) is given by
\begin{equation}
X^5+\kappa \tau_0 X^4 -X^3 (1+\kappa^2+\beta^2 \kappa^2)-X^2 \kappa \tau_0 (1+\kappa^2)+X \kappa^2(\beta^2(1+\kappa^2)-\tau_2)-\kappa^3 \tau_0 \tau_2 =0,	
\label{dispersion}
\end{equation}
where $\kappa = c k /\omega_p$ is the rescaled wavelength, and
\begin{eqnarray*}
&& \tau_0 = \frac{4\lambda}{3}\left(\frac{\rho_0}{mc}\right)^{1/3} \left(\frac{T_{02}}{mc^2}\right)^{1/3},\\
&& \tau_2 = \frac{T_{20}}{mc^2},\\
&& \beta = \frac{3\alpha^2 T_{02}}{m^2 c^2 T_{20}}.
\end{eqnarray*}
Here we have fixed a gauge such that $A_x(z,t)=\delta A_x {\rm e}^{i (k z-\omega t)}$. 
The condition for a good choice of parameters $(\alpha,\lambda)$ is that the small-$k$ are unstable. We expand the dispersion relation for small $ck/\omega_p$:
$$
\omega(k)= c k X_0 + O(k^3),
$$ 
where $X_0$ solution of 
$$
X_0^3 +\tau_0 X_0^2 -X_0 (\beta^2-\tau_2)+\tau_0 \tau_2 =0. 
$$
The discriminant of this polynomial should be negative to have unstable modes. This happens if $\tau_0$ is large enough (or equivalently if $\lambda$ is large enough). More precisely, the condition is
$$
8\frac{\tau_0^2}{\tau_2}\geq \bar{\beta} (\bar{\beta}^3+(\bar{\beta}^2+8)^{3/2})-20 \bar{\beta} -8,
$$
where $\bar{\beta}=\beta / \sqrt{\tau_2}$.  We notice that for $\bar{\beta}\leq 1$, all values of $\tau_0\geq 0$ lead to unstable modes for small $c\,k/\omega_p$.  In Fig.~\ref{fig:smallk}, we represent the set of
parameters $(\beta,\tau_0)$ for which these modes are unstable.

\begin{figure}[htb]
\centering
\includegraphics[width=0.9\textwidth]{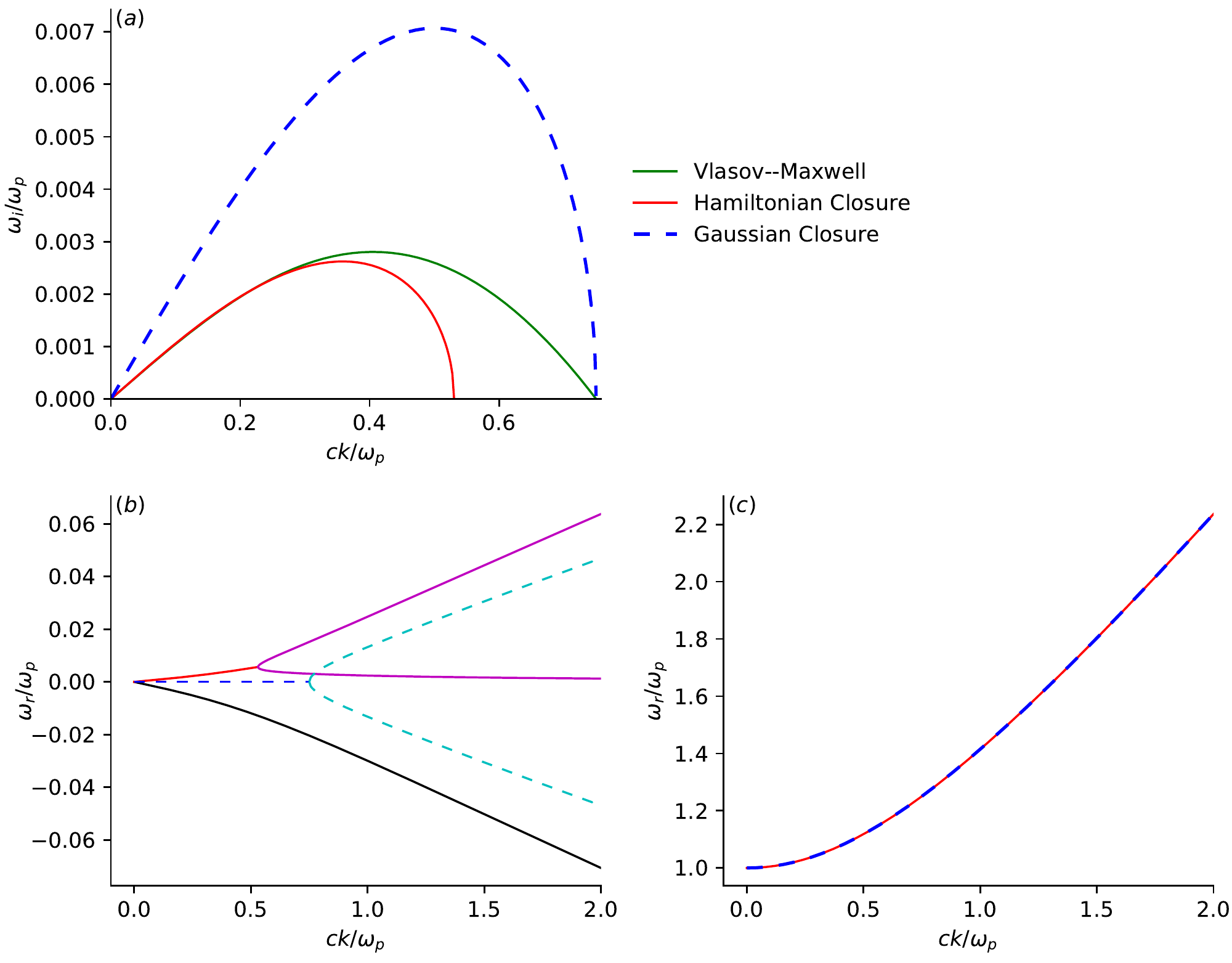}
\caption{\label{fig:dispersion} Solutions of the dispersion relation~(\ref{dispersion}) for $\tau_2=1.25\times 10^{-3}$, $\beta/\sqrt{\tau_2}=1.05$ and $\tau_0/\sqrt{\tau_2}=0.08$. The temperature anisotropy is $\sqrt{T_{20}/T_{02}}=1.25$.}
\end{figure}

Figure \ref{fig:dispersion} shows the solutions of Eq.~(\ref{dispersion}) compared with solutions of the corresponding dispersion relation for the Gaussian closure and to discrete mode of the linearized
Vlasov--Maxwell system.  The real and imaginary parts of the solution corresponding to the unstable mode are shown in panels $(a)$ and $(b)$, respectively for the Hamiltonian (red) and Gaussian
(blue) closures.  The growth rate of the unstable mode of the linearized Vlasov-Maxwell system (green) is purely imaginary. The parameters are $\tau_2=1.25\times 10^{-3}$, $\beta/\sqrt{\tau_2}=1.05$, $\tau_0/\sqrt{\tau_2}=0.08$ and a temperature anisotropy $\sqrt{T_{20}/T_{02}}=1.25$. In terms of the parameters of the Hamiltonian fluid closure, these parameters give 
\begin{eqnarray*}
&& \frac{\alpha}{mc}\approx 0.14,\\
&& \lambda \left(\frac{\rho_0}{mc} \right)^{1/3}\approx 2.29\times 10^{-3}. 
\end{eqnarray*}
The first equation is compatible with the assumption that the model is non-relativistic in the whole range of $P_x$ where the Hamiltonian model is non-singular. 

Panel $(b)$ shows the bifurcation of the unstable mode into a pair of real solutions when $k$ exceeds a critical value for the Hamiltonian (magenta) and Gaussian (cyan) closures.  In addition, panel $(b)$ shows a purely real solution of Eq.~(\ref{dispersion}) (black) that is absent from the Gaussian closure.
Shown in panel $(c)$ is the high-frequency solution to Eq.~(\ref{dispersion}) (red) and the corresponding solution of the dispersion relation of the Gaussian closure (blue).  The Vlasov-Maxwell system has a very weakly damped solution with a nearly identical value of $\omega_r$ (not shown).

\section{Conclusion}
We derived Hamiltonian fluid models based on the closure of the 1.5D Vlasov-Maxwell equations. These reduced models were obtained by solving the Jacobi identity for the closure functions. They possess two key ingredients: First, they depend on some parameters which allow the fluid model to be adjusted to reproduce quantitatively some features of the corresponding kinetic equations. Second, they exhibit singularities which indicate a region of phase space where the fluid reduction fails. We notice that another way to derive the closures for the reduced fluid model to be Hamiltonian is to require that the fluid model possesses a Casimir invariant of the entropy-form. This generalizes a result found in the one-dimensional case~\cite{Perin14}. \\
For a wide range of parameters, the Hamiltonian fluid models exhibit unstable transverse electromagnetic modes in the presence of anisotropy in the velocity distribution, in agreement with the kinetic model.

\section*{Acknowledgments}
This material is based upon work supported by the National Science Foundation under Grant No.  DMS-1440140 while the authors were in residence at the Mathematical Sciences Research Institute in Berkeley, California, during the Fall 2018 semester.  This work has been carried out within the framework of the French Federation for Magnetic Fusion Studies (FR-FCM).  BAS was supported in part by the National Science Foundation under Contract No.\ PHY-1535678.

\end{document}